# *Residual gas absorption effect on the electronic structure of Cr-doped Bi$_2$Se$_3$*

**T Yilmaz[1], W Hines[2], S Alraddadi[3], J I Budnick[2] and B Sinkovic[2]**

[1]Department of Physics, Science and Literature Faculty, Uludag University, Bursa 16059, Turkey
[2]Department of Physics, University of Connecticut, Storrs, Connecticut 06269, United States
[3]Department of Physics, Umm Al-Qura University, Makkah 24382, Kingdom of Saudi Arabia

**ABSTRACT:** In this report, we identify the origin of the temperature dependence of the surface energy gap in impurity-doped topological insulators. The gap at the Dirac point and its variation with temperature were studied by using angle-resolved photoemission spectroscopy in Cr-doped Bi$_2$Se$_3$. Our valence band photoemission results revealed that the gap varies with temperature due to residual gas condensation on the sample surface with cooling. Adsorbate on the surface of the sample creates an electron doping effect that modifies the chemical potential of the system resulting in the change of the gap with variable temperature. Such electron doping can weaken the ferromagnetism and lead to a bulk band contribution in the transport measurements. Also, a larger energy gap is required to suppress the thermal excitations for the quantum anomalous Hall effect. Therefore, such effects can hinder the quantum anomalous Hall state at higher temperatures. Resolving this issue can pave the way for enhancing the observation temperature of the quantum anomalous Hall effect. Therefore, our findings can play a significant role in the discovery of the high-temperature quantum anomalous Hall effect in impurity-doped topological insulators.

## 1. Introduction

Topological insulators (TIs) attract scientists due to their distinct electronic structure and the potential to revolutionize information technology. They are insulators in the bulk but possess very special metallic surfaces due to strong spin-orbit coupling [1-4]. Surface electrons of TIs obey the massless Dirac equation that the electron energy disperses linearly with respect to its momentum. This forms the Dirac cone in momentum space with a special crossover singularity known as the Dirac point (DP) [4]. On the surface of TIs, time reversal symmetry (TRS) constrains the momentum and spin of electrons locked perpendicularly to each other leading the absence of backscattering. Hence, surface states of TIs are topologically protected from non-magnetic impurity scattering and sample imperfections [4-6].

In contrast, a long-range magnetic order can open a surface energy gap at the DP by breaking time reversal symmetry (TRS) [7-9]. The gap has been experimentally observed in various transition metal doped Bi$_2$Se$_3$ topological insulators by angle-resolved photoemission spectroscopy (ARPES) [9-12]. Magnetic TIs with broken TRS are of great interest for being a new avenue to observe the quantum anomalous Hall effect (QAHE) [13, 33-35].

Growing interest in magnetic TIs, however, has brought about several unexpected results. For example, recent studies on Cr and Mn-doped $Bi_2Se_3$ revealed that the Dirac cone can be gapped without a need for magnetization [14-18, 21]. This gap at the DP was attributed to scattering of Dirac electrons by the impurity induced states located at and around the DP [19-21]. In addition, the gap size in $Bi_{2-x}Cr_xSe_3$ (for x > 0) depends on the chemical potential that it can be controlled by Mg incorporated into the bulk of the sample [14].

Another intriguing realization is the temperature dependence of the gap [14, 23]. For example, the 70 meV gap measured at room temperature (RT) for $Bi_{1.96}Cr_{0.04}Se_3$ sample almost vanishes at 80 K [14]. The physics behind the temperature dependence of the gap has not been investigated and still is awaiting to be answered. Here, we aim to solve this puzzle in Cr-doped $Bi_2Se_3$ ($Bi_{2-x}Cr_xSe_3$) by performing temperature dependent ARPES, x-ray photoemission spectroscopy (XPS), and x-ray diffraction (XRD) experiments. We found that the temperature dependence of the gap arises from the residual gas condensation on the sample surface with sample cooling. The residual gas dopes the system with electrons and modifies the chemical potential of the system resulting in the suppression of the gap at low temperature. The findings in this report can contribute the understanding of why the QAHE can be observed only at extremely low temperature in other systems.

## 2. Experiment

Molecular beam epitaxy (MBE) is employed for growing high quality pristine and Cr-doped $Bi_2Se_3$ thin films on $Al_2O_3$ (0001) substrates. The two-step method is applied by depositing a few-quintuple layers (QL) of the film at 390 K and deposited more at 490 K substrate temperature. The samples were annealed for 10 minutes at 490 K to promote the surface quality of the films. Angle resolved photoemission spectroscopy and valence band spectra were collected with He I$\alpha$ radiation (hv = 21.2 eV) and Scienta SES-100 electron analyzer. Core levels were recorded with Al K$\alpha$ x-rays (hv = 1486.6 eV). The MBE growth chamber with a base pressure of 1x10-9 torr and the photoemission chamber with a base pressure of 2x10-10 torr were vacuum-interlocked. Hence, photoemission measurements were carried without breaking the vacuum. This experimental setup provides a unique way of monitoring the electronic structure of freshly grown samples.

## 3. Results and discussions

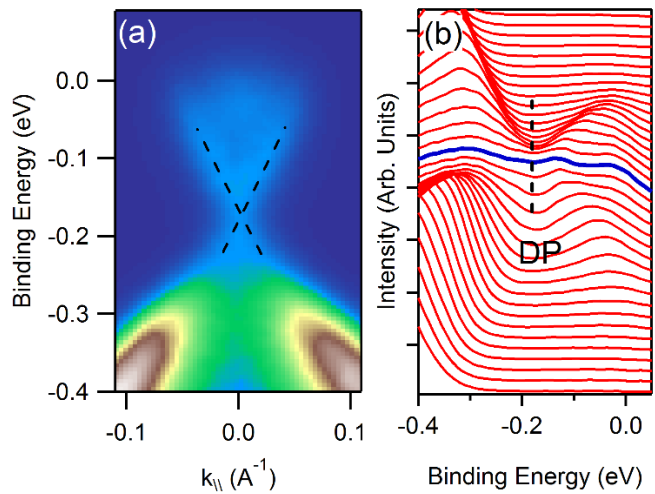

**Figure 1.** (a) ARPES energy mapping of $Bi_2Se_3$ (x = 0) at RT and (b) corresponding energy dispersive curves (EDC). Dashed black lines in (a) represent the topological surface states of Bi2Se3. Spectrum was recorded with 21.2 eV (He I$\alpha$) photon energy along the $\Gamma$-M direction.

The unique electronic structure of $Bi_2Se_3$ and the impact of Cr-doping on the surface states of $Bi_2Se_3$ were investigated by means of ARPES. figure 1(a) shows the ARPES spectrum of a pristine $Bi_2Se_3$ film at RT. Gapless and linear dispersion of the surface states constructing the DP at Γ (000) and 180 meV binding energy demonstrate the topological insulator character of the $Bi_2Se_3$ film. In In addition, corresponding energy dispersive curves (EDCs) are presented in figure 1(b) in which the gapless feature of the topological surface states is marked with a solid blue line.

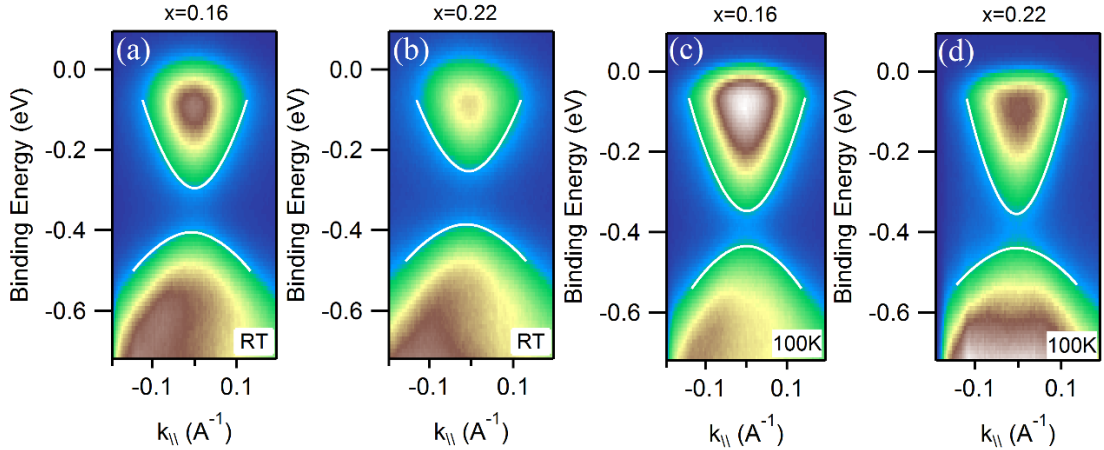

**Figure 2.** (a) and (b) are ARPES data at RT for $Bi_{1.84}Cr_{0.16}Se_3$ (x = 0.16) and $Bi_{1.78}Cr_{0.22}Se_3$ (x = 0.22), respectively. (c) and (d) are ARPES energy maps at 100 K for x = 0.16 and x = 0.22, respectively. Spectra were recorded with 21.2 eV (He Iα) photon energy along the Γ-M direction. The surface energy gap is measured from the bottom of the upper surface state to the top of the lower surface state as indicated with solid white lines.

Cr-doping has a sudden and dramatic impact on the electronic structure of $Bi_2Se_3$. First, Cr incorporating into the bulk of $Bi_2Se_3$ causes broader energy bands that limit the ability to distinguish the surface and the bulk states. Second, it destroys the Dirac-like surface states and opens an energy gap at DP as shown in figure 2(a) for the $Bi_{1.84}Cr_{0.16}Se_3$ film. Such gap opening at DP indicates the mass acquisition of the Dirac fermions by breaking TRS and lifting the Kramer's degeneracy. However, the ferromagnetic transition temperature was reported to be 35 K for Cr-doped $Bi_2Se_3$ [16]. Therefore, the gap is not expected to be resolved at RT. In addition, increasing Cr content from x = 0.16 to 0.22 results in a larger energy gap (figure 2(b)). The gap is measured from the bottom of the upper surface states to the top of the lower sur-face states as indicated with white solid parabolas in figures 2(a) and 2(b), and it is measured to be 100 meV for x = 0.16 and 130 meV for x = 0.22 at RT.

Furthermore, we recorded the electronic structure of $Bi_{1.84}Cr_{0.16}Se_3$ and $Bi_{1.78}Cr_{0.22}Se_3$ at 100 K (figures 2(c) and 2(d)). The gap shows temperature dependence and it is measured to be 75 meV at 100 K for both samples. Since Cr-doped $Bi_2Se_3$ is not magnetic above the 35 K [16], the gap is not expected to fluctuate from RT to 100 K. In addition, measuring the same gap size for different Cr content at low temperature suggests that the gap size is Cr content independent at 100 K or there are some other factors merging with the decreasing temperature.

A recent work also pointed out that the gap magnitude changes with respect to temperature in $TlBi(S_{1-x}Se_x)_2$ samples, which is a TI for x > 0.6 [23]. Increasing the S content above x = 0.6 opens a gap at the DP that shows similar temperature dependence as observed in our samples. $TlBi(S_{1-x}Se_x)_2$ is a non-magnetic material for all x values. Thus, the dependence of the gap on temperature is attributed to fluctuation or randomness, which can be suppressed at low temperature [23].

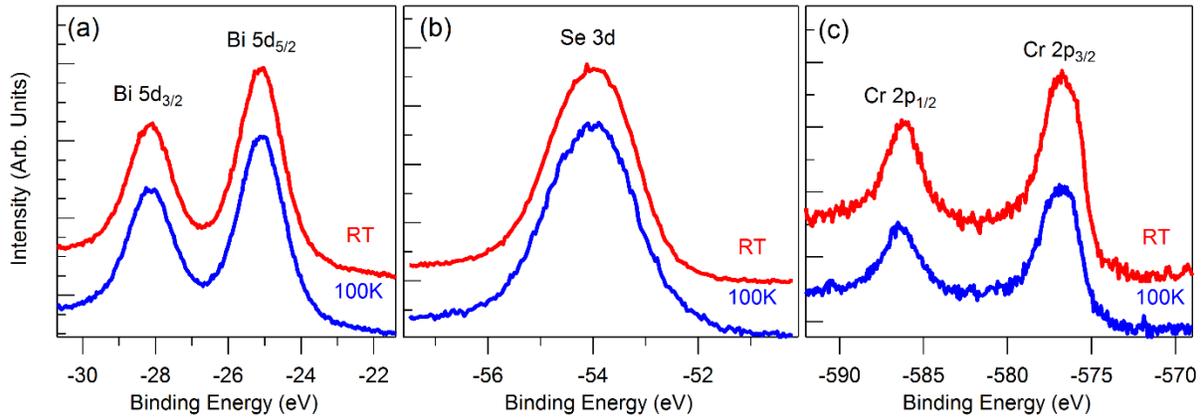

**Figure 3.** XPS core level peaks of (a) Bi 5d, (b) Se 3d and (c) Cr 2p obtained from $Bi_{1.84}Cr_{0.16}Se_3$ sample at RT with Al Kα (1458.6 eV) source.

There could be several driving factors behind the temperature dependence of the gap. We start our argument by performing a XPS study. Theoretical findings showed that transition metal substitution only occurs with Bi in $Bi_2Se_3$ and results in opening an energy gap at the DP [22]. Therefore, the temperature dependence of the gap may originate from the evolution of the chemical environment with variable temperature. Thus, we recorded the Bi 5d, Se 3d and Cr 2p peaks at RT and 100 K for the $Bi_{1.84}Cr_{0.16}Se_3$ sample (figure 3). The core-level peaks do not show any prominent difference between the two temperatures except for 70 meV shift to the higher binding energy with cooling. In addition, at 100 K, Cr 2p1/2 and 2p3/2 have the binding energy of -586.2 eV and -576.4 eV, respectively. These binding energies correspond to +3 oxidation states indicating that Cr is still in the Bi-lattice site at 100 K. Hence, the chemical structure of $Bi_{1.84}Cr_{0.16}Se_3$ is stable with temperature confirming that variation in the chemical environment is not a possible link between the gap size and temperature.

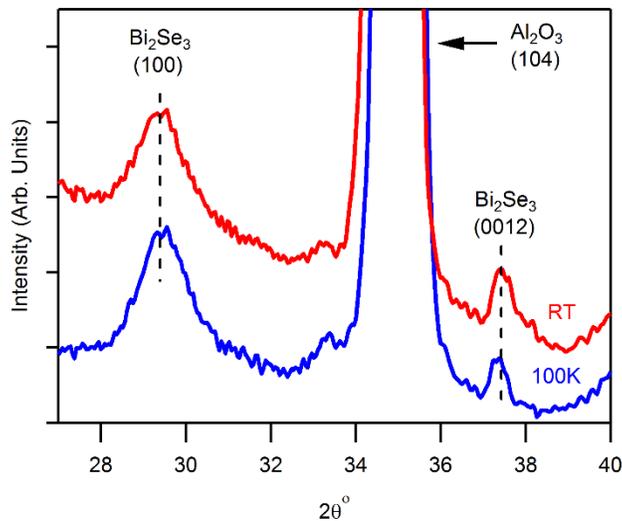

**Figure 4.** X-ray diffraction patterns obtained RT (red) and 100 K (blue) Bi1.84Cr0.16Se3 films grown on $Al_2O_3$ (0001) substrates. Data was collected with Cu Kα radiation (λ = 1.5418 Å). Dashed lines are for guiding eyes.

The evolution of the gap with temperature could also arise from the expansion or compression of the c-lattice parameter. A recent work revealed that the gap opened at the DP of $Bi_2Se_3$ can be controlled by strain applied along the surface normal [32]. To investigate if this is the case in our samples, we performed XRD on $Bi_{1.84}Cr_{0.16}Se_3$ sample at RT and 100 K (figure 4). The c-lattice parameter is calculated to be 28.65±0.02 Å at both temperatures indicating the negligible effect of temperature on the c-lattice parameter. Therefore, the strain of c-lattice parameter does not play a role in the change of the gap with temperature.

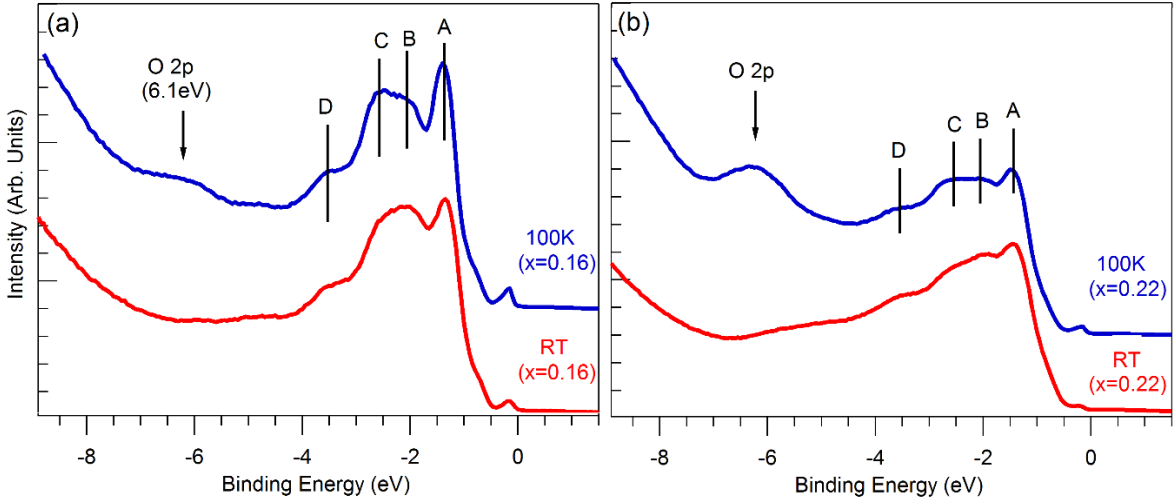

**Figure 5.** (a) and (b) are EDCs obtained from $Bi_{1.84}Cr_{0.16}Se_3$ and $Bi_{1.78}Cr_{0.22}Se_3$ respectively, at RT (red) and 100 K (blue) at Γ (000) with 21.2 eV (He Iα) photon energy. Solid black lines represent the peak positions in energy scale.

Another origin of the temperature dependence of the gap could be the modification of the chemical potential. Therefore, we study the valence states of $Bi_{1.84}Cr_{0.16}Se_3$ and $Bi_{1.78}Cr_{0.22}Se_3$ at RT and 100 K (figures 5(a) and 5(b)). Energy dispersive curves (EDCs) of $Bi_{1.84}Cr_{0.16}Se_3$ recorded at RT consists of only Bi 6p and Se 4p valence band states. We observed one intense peak located at 1.35 eV (A), which is the bonding character of Se 4p. Another prominent peak (B) is located at 2.05 eV (B) dominated by Bi 6p bonding orbit. Other two peaks located at 2.45 eV(C) and 3.5 eV (D) binding energy originate from Bi 6p and Se 4p antibonding orbitals, respectively [24-25]. Similarly, EDC of $Bi_{1.78}Cr_{0.22}Se_3$ also dominated by Se 4p and Bi 6p bonding and antibonding peaks.

At 100 K, however, a peak appears at 6.1 eV marked with an arrow in figures 5(a) and 5(b). The binding energy of the peak corresponds to an oxygen 2p [26-28]. This arises from the residual gas ($H_2$, CO, $CO_2$, and $H_2O$) condensation on the sample surface at low temperature. The condensation of the residual gas has an electron doping effect on the electronic structure of TIs [29-31, 40], which is consistent with the shift of our EDCs and XPS peaks to the higher binding energy. Hence, our data confirm the residual gas condensation on the sample surface with cooling. This explains the mechanism behind the temperature dependence of the gap. Such electron doping modifies the chemical potential of the system and consequently the size of the gap as in the case of Mg doping effect on the electronic structure of $Bi_{2-x}Cr_xSe_3$ (for x > 0) [14].

For further confirmation, we imitate the electron doping effect by growing extra Cr metal on the surface of $Bi_{1.78}Cr_{0.22}Se_3$ (figure 6). Valence band spectrum of the undoped sample shifts to higher binding energy upon Cr deposition confirming the more electron-rich surface (figure 6(c)). As expected, the gap shrinks from 130 meV to 80 meV as a result of electron doping effect. Since the spectra were taken at RT before and after the surface deposition of Cr, EDCs do not show any presence of residual gas on the sample sur-face. Hence, we provide concrete results to confirm that gap exhibits temperature dependence due to the modified electronic structure by electron doping of residual gas condensed on the sample surface.

Furthermore, the correlation between the size of the gap and the Fermi level can be understood as follows. The gap in the non-magnetic state is attributed to scattering of the Dirac electrons by the strong impurity resonance states located at and around the DP [19-21]. This can open up a gap at the DP as large as 100 meV, which is compatible with the gap size obtained from our ARPES measurements. The influence of the resonance states depends on the binding energy of the DP and it diminishes with larger energy separation between the DP and the Fermi level [38-39]. As a consequence of the residual gas condensation on the sample surface, the energy bands shift to higher binding energy. Hence, the size of the gap can shrink as a result of weakened strength of the resonance states.

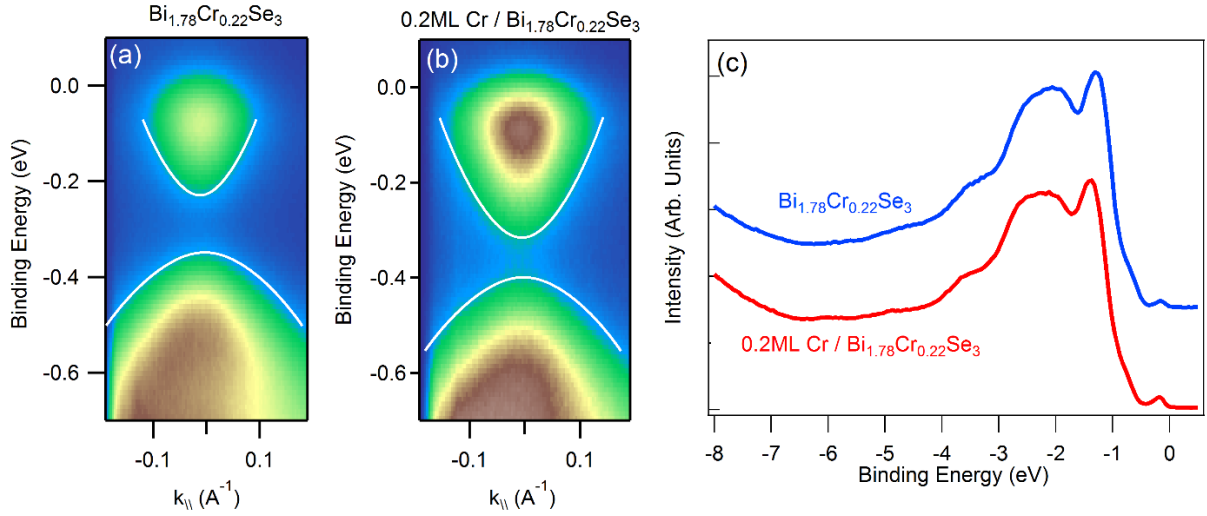

**Figure 6.** (a) and (b) ARPES spectra of $Bi_{1.78}Cr_{0.22}Se_3$ and 0.2 ML Cr covered $Bi_{1.78}Cr_{0.22}Se_3$, respectively. (c) EDCs taken at Γ (000) with 21.2 eV (He Iα) photon energy. The spectra were collected at RT.

## 4. Conclusion

In conclusion, we present temperature dependent photoemission and XRD experimental results obtained from $Bi_{1.84}Cr_{0.16}Se_3$ and $Bi_{1.78}Cr_{0.22}Se_3$ film samples. We found that residual gas condensation on the sample surface with lowering temperature is responsible for the temperature dependence of the gap. Our findings can have a significant importance for enhancing the observation temperature of the QAHE, which is at the center of interest in the TI research area. It has been observed only in Cr and V-doped $(Bi_xSb_{1-x})_2Te_3$ in the millikelvin range [33-34]. One of the essential conditions of the QAHE in TIs is to locate the Fermi level within the bulk band gap to obtain the insulating bulk state and achieve the conduction only by the edge states in transport measurements. Therefore, Bi is added in the $Sb_2Te_3$ topological insulator to tune the Fermi level inside the bulk band gap where the DP is located [35]. However, our spectroscopic studies revealed that air condensation on the sample surface behaves in contrast by pushing bulk band gap away from the Fermi level, which can obstruct the observation of the QAHE. In addition, early studies showed that the formation of $BiO_x$ layers on the sample surface in the residual gas affects the electrical properties and the chemical environment of the $Bi_2Se_3$ [36-37]. Such formation of the oxide layers on the sample sur-face bury the topological material under a surface layer affecting the transport measurement. As a result, such dramatic modification on the electronic structure and transport properties can play an important role in the absence of the QAHE in Cr-doped $Bi_2Se_3$ and the observation of the QAHE at only in the millikelvin range in the other systems.

Another problem arising due to the modification of the electronic structure is to affect the magnetic properties of the material. In magnetic TIs, TRS is expected to be broken due to the Ruderman-Kittel-Kasuya-Yosida (RKKY) type of ferromagnetism that forms if the gap is near the Fermi level [9]. However, our spectroscopic study shows that a shift of the energy bands to higher binding energy which could prevent the formation of RKKY type of ferromagnetism.

These problems, indeed, could be overcome by capping the films before taking the samples out from the ultra-high vacuum chamber for magnetic and transport measurements. Te capping has been widely used for such purposes. However, it is claimed that it has an unfavorable effect on the transport measurements [34]. Therefore, better protection of the surface quality with a capping layer without changing the sample properties may be required in order to achieve the high-temperature QAHE.

**Acknowledge**
This work was funded by the University of Connecticut under the UCONN-REP (Grant No. 4626510).

**References**
   [1]   Kane C L Kane and Mele E J 2005 Phys. Rev. Lett. 95 226801
   [2]   Bernevig B A and Zhang S-C, 2006 Phys. Rev. Lett. 96 106802
   [3]   Moore J E and Balents L, 2007 Phys. Rev. B 75 121306
   [4]   Fu L, Kane C L and Mele E J 2007 Phys. Rev. Lett. 98 106803
   [5]   Murakami S. 2007 New J. Phys. 9 356
   [6]   M. Z. Hasan and C. L. Kane 2010 Rev. Mod. Phys. 82 3045
   [7]   Bernevig B A, Hughes T L and Zhang S.-C 2006 Science 314 1757
   [8]   Liu Q, Liu C-X, Xu C, Qi X.-L and Zhang S-C 2009 Phys. Rev. Lett. 102 156603
   [9]   Chen Y L et al 2010 Science 329 659
 [10]   Liu M et al 2012 Phys. Rev. Lett. 108 036805
 [11]   Xu S-Y et al 2012 Nat. Phys. 8 616
 [12]   Wray L A et al 2011 Nature Phys. 7 32
 [13]   Chang C-Z et al 2013 Science 340 167
 [14]   Chang C-Z et al 2014 Phys. Rev. Lett. 112 056801
 [15]   Liu W et al 2015 ACS Nano 9 10237
 [16]   Kou X F et al 2012 J. Appl. Phys. 112 063912
 [17]   Haazen P P et al 2012 Appl. Phys. Lett. 100 082404
 [18]   Collins-McIntyre L J et al 2014 Europhys. Lett. 107 57009
 [19]   Black-Schaffer A M and Balatsky A V 2012 Phys. Rev. B 85 121103(R)
 [20]   Black-Schaffer A M and Balatsky A V 2012 Phys. Rev. B 86 115433
 [21]   Sanchez-Barriga J et al 2016 Nat. Commun. 7 10559
 [22]   Abdalla L B, Seixas L, Schmidt T M, Miwa R H and Fazzio A 2013 Phys. Rev. B 88 045312
 [23]   Sato T et al 2011 Nat. Phys. 7 840
 [24]   Nascimento V B et al 1999 J. Electron Spectrosc. Relat. Phenom. 104 99
 [25]   Takahashi T, Sagawa T and Hamanaka H 1984 J. Non-Cryst. Solids 65 261
 [26]   Wertheim G K and Hüfner S 1972 Phys. Rev. Lett. 28 1028
 [27]   Raaen S and Murgai V. 1987 Phys. Rev. B 36 887
 [28]   Simonson R J, Wang J R and Ceyer S T 1987 J. Phys. Chem. 91 5681
 [29]   Benia H M, Lin C, Kern K and Ast C R 2011 Phys. Rev. Lett. 107 177602
 [30]   King P D C et al 2011 P. Phys. Rev. Lett. 107 096802
 [31]   Durand C et al 2016 Nano Lett. 16 2213

[32] Liu W, Peng X, Tang C, Sun L, Zhang K and Zhong J 2011 Phys. Rev. B 84 245105
[33] Chang C-Z et al 2013 Science 340 6129
[34] Chang C-Z et al 2015 Nat. Mater. 14 473
[35] Chang C-Z et al 2013 Adv. Mater. 25 1065
[36] Yeh Y C et al 2016 J. Phys. Chem. C 120 3314
[37] Kong D et al 2011 ACS Nano. 5 4698
[38] Teague M L, Xiu F-X, He L, Wang K.-L and Yeh N-C 2012 Solid State Commun. 152 747
[39] Zheng S H et al 2015 Phys. Lett. A 379 2890
[40] Bahramy M S et al 2012 Nat. Commun. 3 1159